\documentstyle{elsart}
\begin{document}
\begin{frontmatter}
\title{Gluonic Correlations in Matter}

\collab{Su Houng Lee$^1$  and Ismail Zahed$^{2,3}$ }
\address{${}^1$ Department of Physics and Institute of Physics and Applied Physics,\\ 
Yonsei University, Seoul 120-749, Korea \\
${}^2$  School of Physics, Korea Institute for Advanced Study, Seoul 130-012, Korea\\
${}^3$  Department of Physics and Astronomy, State University of New York,\\
Stony Brook, NY 11794-3800, USA  }

\begin{abstract}

We derive the analogue of the QCD low energy theorems for the scalar and pseudoscalar
gluonic correlators in nuclear matter. We find that the scalar correlations are depleted
while the pseudoscalar correlations are enhanced to leading order in the nuclear matter 
density. We briefly discuss the consequences of these findings on the QCD spectrum.

\end{abstract}

}

\end{frontmatter}

\section{INTRODUCTION}

QCD Low energy theorems (LET) for the gluonic correlation functions have provided insights 
into the QCD dynamics both in the vacuum and in the nuclear medium.
Indeed, the LET  for the pseudoscalar gluon correlation function provides a powerful
constraint on the contribution of the $\eta'$ mass~\cite{Witten79,Veneziano79} and the
vacuum gluon condensate~\cite{NSVZ81}. We recall that a quantitative calculation of glueball masses 
using QCD sum rule analysis\cite{SVZ,qsr}, require the use of LET to provide the necessary
subtraction constants needed in the dispersion 
analysis\cite{qsr,Narison98}.   Recently, the extension of the LET
to finite temperature and density have  been 
derived for the correlation functions of the scalar gluon 
operators\cite{kap}.   These theorems provide strong constraints
to be satisfied by lattice calculations or any effective model 
calculations of QCD in matter.    Here, we will derive anew the LET
for the scalar gluon correlation functions at low density and show
the equivalence of our result to that of ref.\cite{kap}.  
Then, we will derive similar LET for the pseudoscalar gluonic 
correlation functions in two ways : First, by assuming that the self-dual gauge
configurations are dominant in vacuum and (dilute) matter; Second,
by using a generic heavy-quark expansion. In both cases, we will derive
general relations on the density dependence of the correlation function 
in terms of differential operators. We briefly comment on the relevance of our
results to the QCD spectrum.

\section{LET in vacuum: review}

Let us start with the definition of the scalar and pseudo scalar gluonic 
two point function.

\begin{eqnarray}
\label{def}
S(Q^2)=i\int d^4x e^{iqx} \langle T^* \frac{3 \alpha_s}{4 \pi}G^2(x) 
 \frac{3 \alpha_s}{4 \pi}G^2(0) \rangle  \nonumber \\
P(Q^2)=i\int d^4x e^{iqx} \langle T^* \frac{3 \alpha_s}{4 \pi}G \tilde{G}(x) 
 \frac{3 \alpha_s}{4 \pi}G \tilde{G}(0) \rangle .
\end{eqnarray}
Here, $G^2=G_{\mu \nu}^a G_{\mu \nu}^a$ and 
 $G \tilde{G}=G_{\mu \nu}^a \frac{1}{2} \epsilon_{\mu\nu \alpha \beta}
 G_{\alpha  \beta}^a$.  
The low energy theorem for the scalar gluon operator follows from the 
formula\cite{NSVZ81},
\begin{eqnarray}
\label{svz1}
{d \over  d(-1/4g_0^2)} \langle O \rangle
= i\int d^4x  \langle T^*\,O (x) 
 g_0^2G^2(0) \, \rangle ,
\end{eqnarray}
where $g_0$ is the bare coupling constant of QCD.  Now using renormalization 
group arguments, one notes that the bare coupling is related to the ultraviolet 
cut-off $M_0$ via,
\begin{eqnarray}
\label{rg}
\langle O \rangle ={\rm const}
 \left[ M_0 exp(- \frac{8\pi^2}{bg_0^2}) \right]^d,
\end{eqnarray}
with $b=11-\frac{2}{3}N_f$ to one-loop.  
Therefore, the left hand side of  eq.(\ref{svz1}) yields,
\begin{eqnarray}
\label{svz2}
{d \over  d(-1/4g_0^2)} \langle O \rangle
= d \frac{32 \pi^2}{b} \langle O \rangle 
\end{eqnarray}
Substituting $O=\frac{3 \alpha_s}{4 \pi}G^2(0)$ into eq.(\ref{svz1}) and
using eq.(\ref{svz2}) we find the LET for the scalar gluon 
correlation function.
\begin{eqnarray}
\label{LET1}
S(Q^2 = 0)= \frac{18}{b} \langle \frac{\alpha_s}{\pi} 
G^2 \rangle
\end{eqnarray}

The situation is more subtle for the pseudoscalar gluonic correlation
function.  Indeed,  in the presence of light quarks, 
the pseudoscalar gluon field is a total differential of the 
light quark axial-current, i.e.  $\frac{3 \alpha_s}{4 \pi}G \tilde{G}= \sum_q 
\partial_\mu 
\bar{q} \gamma_\mu \gamma^5 q$ ($m_q\rightarrow 0$). Hence,
\begin{eqnarray}
P(Q^2=0) =0\,\,.
\end{eqnarray}
This is not true in the absence of light quarks (NLQ).  The approximate 
LET for this case was originally derived in ref.\cite{NSVZ81} by 
assuming that self-dual field configurations dominate the functional 
integral\cite{NSVZ81} over the gauge fields.  
To show this, consider the QCD
functional integral with nonzero $\theta$,  
\begin{eqnarray}
\label{func}
Z=\int [dA] exp ( i I)
\end{eqnarray}
where
\begin{eqnarray}
I=\int d^4x \left[ -\frac{1}{4}G^2 -\theta \frac{g_0^2}{32\pi^2}
G \tilde{G} \right]
\end{eqnarray}
Differentiating eq.(\ref{func}) twice with respect to 
 $\theta$, we have,
\begin{eqnarray}
\label{ps1}
{\partial^2 \epsilon \over \partial \theta^2}
=-i \int d^4x \langle T^*\, \frac{g_0^2}{32\pi^2}
G \tilde{G}(x),\frac{g_0^2}{32\pi^2}
G \tilde{G}(0) \, \rangle 
\end{eqnarray}
where $\epsilon$ is the vacuum energy density.   
To evaluate the left hand side of eq.(\ref{ps1}), we note 
that the energy density is related to the trace of the energy 
momentum tensor $\theta_{\mu \nu}$,
\begin{eqnarray}
\label{emt}
\epsilon=\frac{1}{4} \langle \theta_\mu^\mu \rangle
=\langle \frac{\beta(\alpha_s)}{16 \alpha_s}G^2 \rangle 
\stackrel{1-loop}{\approx} \langle \frac{-b\alpha_s}{32 \pi}G^2 \rangle\,\,.
\end{eqnarray}
Now we assume that the self-dual (anti-self dual) field dominates
the functional integral.  This changes the euclidean action 
as follows,
\begin{eqnarray}
\label{sd}
I_E= i \int d^4 x \left[ -\frac{1}{4g_0^2}\bar{G}^2 
-\theta \frac{i}{32\pi^2} \bar{G} \bar{\tilde{G}} \right]
\rightarrow
i \int d^4 x \left[ -\frac{1}{4}(\frac{1}{g_0^2}
+\frac{i \theta}{8 \pi^2} )\bar{G}^2 \right]
\end{eqnarray}
where $\bar{G}$ is obtained by redefining the gauge field 
 $A_\mu \rightarrow \frac{1}{g_0} \bar{A}_\mu$.
 Eq.(\ref{sd}) implies that within this approximation, 
the ultraviolet cutoff $M_0$ is related to $\theta$ 
in a physical quantity by a simple replacement of eq.(\ref{rg}),
\begin{eqnarray}
\label{trg}
\langle O \rangle ={\rm const}
\left[ M_0 exp(- \frac{8\pi^2}{bg_0^2}-\frac{i\theta}{b}
) \right]^d\,\,,
\end{eqnarray}
for small $\theta$~\footnote{We note that for large $\theta$, $2\pi$
periodicity is recovered through the branch-structure of $\epsilon$.}.
With eq.(\ref{trg}) and eq.(\ref{emt}), it is straightforward to calculate the 
left hand side of eq.(\ref{ps1}). We obtain the LET 
for the pseudoscalar gluonic correlations\cite{NSVZ81},
\begin{eqnarray}
\label{LET2}
P_{NLQ}^{sd}(Q^2=0)=-S(Q^2=0)=-\frac{18}{b} \langle \frac{\alpha_s}{\pi}
G^2 \rangle,
\end{eqnarray}
where the superscript $sd$ refers to the fact that it is 
based on the self-dual assumption detailed above. 

We recall that the difference between the 
case with light quarks and without light quarks is the contribution
due to the $\eta'$\cite{Witten79,Veneziano79,NSVZ81}, which adds up to 
give the required zero, i.e.
\begin{eqnarray}
\label{phen1}
P(Q^2=0)=P_{NLQ}(Q^2=0) +
 {| \langle \frac{3 \alpha_s}{4 \pi}  G \tilde{G} |\eta'
 \rangle|^2 \over m_{\eta'}^2 } =0.
\end{eqnarray}
One should note that similar LET  for higher point 
gluonic correlations can be 
obtained by taking further derivatives either with respect to 
 $\theta$ or $-1/4g_0^2$.

\section{LET for the pseudoscalar gluonic current: Heavy Quark 
Mass expansion}

The above derivation for the pseudo scalar current was obtained 
within the self-dual approximation in the path integral.  This approximation
can be averted. Indeed, let us consider a regularization scheme 
with $\theta \neq 0$ in the presence of a heavy quark with mass $m_Q$.   
Substituting the pseudoscalar gluon field for the 
operator $O$ in eq.(\ref{svz1}), we have, 
\begin{eqnarray}
\label{ps2}
{d \over  d(-1/4g_0^2)} \langle \frac{3 \alpha_s}{4 \pi}G \tilde{G} \rangle
= i\int d^4x  \langle T^*\, \frac{3 \alpha_s}{4 \pi}G \tilde{G} (x) 
 g_0^2G^2(0) \, \rangle \,\,.
\end{eqnarray}
Now, we will make use of the heavy quark mass expansion\cite{Gen84},  
\begin{eqnarray}
\label{heavyexpansion}
m_Q \bar{Q} i \gamma_5 Q & = & - \frac{\alpha_s}{8\pi} G \tilde{G}+
O \left( \frac{1}{m_Q^2} \right) \nonumber \\
m_Q \bar{Q}  Q & = & - \frac{\alpha_s}{12\pi} G^2+
O \left( \frac{1}{m_Q^2} \right)\,\,.
\end{eqnarray}
Substituting eq.(\ref{heavyexpansion}) into eq.(\ref{ps2}) we obtain,
\begin{eqnarray}
\label{ps3}
{d \over  d(-1/4g_0^2)} \langle 6 m_Q \bar{Q} i \gamma_5 Q \rangle
= i\int d^4x  \langle T^*\, \frac{3 \alpha_s}{4 \pi}G \tilde{G} (x) 
 48 \pi^2 m_Q \bar{Q} Q (0) \, \rangle \,\,.
\end{eqnarray}
In general, such substitution is not allowed when the right hand side 
of eq.(\ref{ps3}) has an $e^{iqx}$ factor as one is calculating the 
operator product expansion.   However, since we are looking at the 
LET dominated by non-perturbative space-like field 
configuration\cite{NSVZ80}, such substitution is valid.

Consider now making a chiral rotation in $\theta \rightarrow \theta +\frac{\pi}{2}$.  
This will only affect the heavy quark mass 
by $m_Q \rightarrow m_Q e^{i \gamma_5 \pi/2}$. As a result, the
scalar quark condensate turns to the pseudoscalar quark condensate and vice versa.  
Substituting the heavy 
quark operators back to the gluon operators, we obtain the following,
\begin{eqnarray}
\label{ps4}
-\frac{2}{32 \pi^2} {d \over  d(-1/4g_0^2)} 
\langle \frac{\alpha_s}{\pi}G^2  \rangle
= i\int d^4x  \langle T^*\, \frac{3 \alpha_s}{4 \pi}G \tilde{G} (x) 
 \frac{3 \alpha_s}{4 \pi}G \tilde{G}(0) \, \rangle \,\,.
\end{eqnarray}
Making use of eq.(\ref{svz2}), yields
\begin{eqnarray}
\label{LET3}
P_{NLQ}^{hq}(Q^2=0)=-\frac{8}{b} \langle \frac{\alpha_s}{\pi}
G^2 \rangle\,\,,
\end{eqnarray}
where, the superscript ${hq}$ refers to the heavy quark mass expansion used.  
This is different from eq.(\ref{LET2}) as obtained within the self-dual 
approximation.   Given the uncertainty in the value of the non perturbative
gluon condensate, both results are consistent with the vacuum phenomenology  
in eq.(\ref{phen1}). 

\section{LET for the scalar gluonic current in nuclear matter}

Here we derive the LET for the scalar gluonic 
current at low density. The same derivation holds for finite temperature.
The starting point is taking the expectation value of eq.(\ref{svz1}) at 
finite density.  Then we make a low density expansion of the left hand side, 
\begin{eqnarray}
\label{lde1}
{ d \over d(-1/4g_0^2)} \langle O \rangle_\rho 
={ d \over d(-1/4g_0^2)} \left( \langle O \rangle_0
+ \rho \langle N| O |N \rangle + {\cal O}(\rho)\right).
\end{eqnarray}
Here, $\langle \cdot \rangle_\rho$ denotes the expectation value at finite 
density, $\rho$ is the nucleon density and $|N \rangle$ a nucleon
state normalized as $\langle N||N \rangle=(2\pi)^3 E_N/m_N \delta^3(0)$. 
This is the linear density approximation.  The derivative of the 
nucleon expectation value is simply obtained by substituting 
 $d\rightarrow d-3$ in eq.(\ref{svz2}).   The derivative of the 
nucleon density can be obtained from the fact that the chemical 
potential $\mu={p_f^2}/{2m_N}$ is an external parameter and is independent of $g_0$.
That is, using
\begin{eqnarray}
\frac{d}{d(-1/4g_0^2)} \left( \frac{p_f^2}{2m_N}
\right)=0
\end{eqnarray}
and 
\begin{eqnarray}
\frac{d}{d(-1/4g_0^2)}m_N 
= \frac{32 \pi^2}{b} m_N,
\end{eqnarray}
we find,
\begin{eqnarray}
\frac{d}{d(-1/4g_0^2)} p_f^2=&&\frac{32 \pi^2}{b} 2m_N\mu, \\
\frac{d}{d(-1/4g_0^2)} \rho=&& \rho \frac{32\pi^2}{b} \frac{3}{2}\,\,.
\end{eqnarray}
Therefore, 
\begin{eqnarray}
\label{lowd1}
{ d \over d(-1/4g_0^2)} \langle O \rangle_\rho
=\frac{32 \pi^2}{b} \left( d \langle O \rangle_0
+ (d-\frac{3}{2})\rho \langle N| O |N \rangle + {\cal O}(\rho)
\right)\,\,.
\end{eqnarray}
 From this, it follows that the changes of the correlator for the
scalar gluonic current to leading order in density is,
\begin{eqnarray}
\Delta S(Q^2)=S(Q^2; \rho)-S(Q^2; \rho=0)
=\frac{45}{4b} \rho \langle N| \frac{\alpha_s}{\pi}
G^2|N \rangle,
\end{eqnarray}
where $S(Q^2,\rho)$ is the correlation function defined in eq.(1) 
calculated at finite density $\rho$.

We note that eq.(\ref{lowd1}) is also consistent 
with the result of ref.\cite{kap} applied to leading order in density.  
In ref.\cite{kap} it was noted that a physical
quantity of mass scale $d$ has the following dependence on 
the temperature or density.
\begin{eqnarray}
\label{dep}
\langle O \rangle_{\rho,T} =\Lambda^d f(\frac{T}{\Lambda},\frac{\mu}{\Lambda}),
\end{eqnarray}
where $\langle  \cdot \rangle_{\rho,T}$ is the expectation value at finite 
density and temperature, and the mass scale $\Lambda$ depends on the coupling 
as in eq.(\ref{rg}) with $d=1$. 
Then, it follows that the differential operator  acting on physical
parameter can be replaced as,
\begin{eqnarray}
\label{diff1}
\frac{d}{d(-1/4g_0^2)}  \langle O \rangle_{\rho,T}
=\frac{32 \pi^2}{b} \left( d-T\frac{\partial}{\partial T}
 - \mu \frac{\partial}{\partial \mu} \right) \langle O \rangle_{\rho,T}.
\end{eqnarray}
When we apply this differential operator to our eq.(\ref{lde1}) and 
note that $\mu \frac{\partial \rho}{\partial \mu}=\frac{3}{2} \rho$,
we recover eq.(\ref{lowd1}).

\section{LET for the pseudo-scalar gluonic current in nuclear medium:
self-dual approximation}

The LET  for the pseudo-scalar gluonic currents at 
low density can be similarly obtained in the self-dual approximation.
Here, we consider the case with no $sea$ light quarks, stressing the effects
of matter on the gauge configurations.
The starting point is the generalization of eq.(\ref{ps1}) to finite density.  
The difference compared to the scalar case is in taking the derivative with respect 
to $\theta$ instead of $g_0^2$.
Again, using the linear density approximation, we have 
\begin{eqnarray}
\label{lde2}
{ d^2 \over d\theta^2} \langle O \rangle_\rho
={ d^2 \over d\theta^2} \left( \langle O \rangle_0
+ \rho \langle N| O |N \rangle + {\cal O}\,(\rho) \right) .
\end{eqnarray}
To evaluate the derivatives with respect to $\theta$, we only need
the relation in eq.(\ref{trg}) and the fact that
${d \mu}/{d \theta}=0$.
To leading order in density, we have
\begin{eqnarray}
\frac{d \rho}{d \theta} & = & \frac{3}{2} (-\frac{i}{b}) \rho 
\nonumber \\
\frac{d^2 \rho}{d \theta^2} & = & \frac{9}{4} (-\frac{i}{b})^2 \rho \,\,.
\end{eqnarray}
Therefore, we have 
\begin{eqnarray}
\label{lde3}
{ d^2 \over d\theta^2} \langle O \rangle_\rho
& = & (\frac{-i}{b})^2 \bigg( d^2 \langle O \rangle_0
+ (\frac{9}{4}+3(d-3)+(d-3)^2)\rho \langle N| O |N \rangle \nonumber \\
& & + {\cal O}\,(\rho) \bigg).
\end{eqnarray}
Substituting this into eq.(\ref{ps1}) gives,
\begin{eqnarray}
\label{lde4}
\Delta P_{NLQ}^{sd}(Q^2=0) & =& P_{NLQ}^{sd}(Q^2=0;\rho)- P_{NLQ}^{sd}(Q^2=0;0)
\nonumber \\
& = & -\frac{225}{32 b}  
\rho \langle N| \frac{\alpha_s}{\pi} G^2 |N \rangle\,\,.
\end{eqnarray}

We can also derive a general formula for eq.(\ref{lde3}) in terms of 
differential operators involving the derivatives with respect to the 
temperature and/or chemical potential.  
Again, we have the $\theta$ dependence of a physical 
parameter through eq.(\ref{dep}) and also eq.(\ref{trg}).  
Therefore, we have
\begin{eqnarray}
\label{linear1}
\frac{d^n}{d\theta^n}  \langle O \rangle_{\rho,T}
=\left( \frac{-i}{b}\right)^n \left( d-T\frac{\partial}{\partial T}
 - \mu \frac{\partial}{\partial \mu} \right)^n \langle O \rangle_{\rho,T},  
\end{eqnarray}
which is consistent with eq.(\ref{lde3}).  
 In the linear density approximation eq.(\ref{linear1}) can be 
written as,
\begin{eqnarray}
\label{linear2}
\frac{d^n}{d\theta^n} ( \langle O \rangle_0+\rho 
\langle N |O |N \rangle)
=\left( \frac{-i}{b}\right)^n \left( d^n \langle O \rangle_0
+ (d-\frac{3}{2})^n 
\rho \langle N |O |N \rangle) \right).
\end{eqnarray} 

\section{LET for the pseudo-scalar gluonic current in nuclear medium:
Heavy quark expansion}

The generalization of the low energy theorem for the pseudo-scalar current
based on the heavy quark expansion  can be obtained in the same way as 
in the scalar case.  The derivation provides for a check on the self-duality
assumption. Starting from the expectation value of eq.(\ref{ps4}) at 
finite density, and substituting eq.(\ref{lowd1}) 
into the left hand side, we obtain,
\begin{eqnarray}
\label{LET4}
\Delta P_{NLQ}^{hq}(Q^2=0) & = & P_{NLQ}^{hq}(Q^2=0;\rho) - P_{NLQ}^{hq}(Q^2=0;0) 
\nonumber \\ &= & -\frac{5}{b} \rho \langle N| \frac{\alpha_s}{\pi}
G^2 |N \rangle.
\end{eqnarray}
Note that the overall factor is different from eq.(\ref{lde4}), but the 
sign is the same.   The relation in terms of differential operators 
can be obtained in the same way as in the scalar case by substituting
eq.(\ref{diff1}) into the left hand side of eq.(\ref{ps4}).

\section{Summary}

We have derived the LET  for scalar and 
pseudoscalar gluonic correlation functions at low 
density.  Although, we have restricted our discussions to 
two point functions, we can easily generalize this method to
higher point functions and to finite temperature.
Since  eq.(\ref{phen1}) is still valid in the medium, we expect,
\begin{eqnarray}
\label{phen2}
\Delta \left( {| \langle \frac{3 \alpha_s}{4 \pi}  G \tilde{G} |\eta'
 \rangle|^2 \over m_{\eta'}^2 } \right) =   - \Delta P_{NLQ}(Q^2=0) <0\,\,.
\end{eqnarray}
The magnitude of the right hand side depends on the approximation
scheme we have used (self-duality or heavy quark).  However, in both cases, the sign is negative.
To leading order in the matter density and
if we were to assume that residues do not change, then 
the $\eta'$ mass will increase, while the scalar glueball mass will decrease.
These observations are worth testing using QCD lattice simulations. The faith
of the $\eta'$ mass in matter may have interesting consequences on low mass
dilepton emissions, pion thresholds as well as the strong CP problem in matter.

\section{acknowledgments}
We thank KIAS for hospitality during the completion of this work.
The work of SHL was supported by  the Brain Korea 21 project and 
by KOSEF grant number 1999-2-111-005-5. The work of IZ was supported
in part by US DOE grant DE-FG02-88ER40388.

\end{document}